\documentclass[twocolumn,showpacs,preprintnumbers,amsmath,amssymb,nofootinbib]{revtex4}

\usepackage{graphicx}
\usepackage{dcolumn}
\usepackage{bm}
\usepackage{graphics}
\usepackage{epsfig}
\usepackage{color}
\usepackage{amssymb}
\usepackage{amsmath}
\usepackage{amsfonts}

\begin{document}
\title{Pressure Corrections to the Equation of State in the
Nuclear Mean Field.}

\author{Jacek Ro\.zynek} \email{rozynek@fuw.edu.pl}
\affiliation{National Centre for Nuclear Research, Ho\.za 69, 00-681 Warsaw, Poland}

\begin{abstract}
We show the connection between stiffness of Equation of State (EoS) in a Relativistic
Mean Field (RMF) of Nuclear Matter (NM) and the existence of a strong violation of
longitudinal Momentum Sum Rule (MSR) in RMF for a finite pressure. The increasing
pressure between nucleons starts to increase the ratio of a nucleon Fermi to average
single particle energy and according to the Hugenholtz-van Hove (HvH) theorem valid
for NM, the MSR is broken in the RMF approach. We propose changes which modify the
nucleon Partonic Distribution Function (PDF) and make (EoS) softer to fulfill MSR sum
rule above a saturation density. The course of EoS in our modified RMF model is very
close to a semi-empirical estimation and to results obtained from extensive DBHF
calculations with a Bonn A potential which produce EoS enough stiff to describe
neutron star properties (mass-radius constraint), especially the mass of "PSR
J1614–2230" the most massive known neutron star, which rules out many soft equations
of state including exotic matter. Other features of the model without free parameters
includes good values of saturation properties including spin-orbit term. An admixture
of additional hyperons are discussed in our approach.
\end{abstract}
\maketitle
\section{Introduction}
Experimentally, in the Deeply Inelastic electron Scattering (DIS) on nuclear targets,
photons with large negative momentum square $-q^2=Q^2>1GeV^2$ and large energy
transfer $\nu$, interacting with partons, probe bound hadrons - a kind of moving
sub-targets. Start with the picture of a nucleus with mass $M_A$ (A - a mass number).
Bj\"orken scaling allows to describe nuclear dynamics by the Structure Function (SF)
$F_2^A(x_A )$ which depends on the Lorentz invariant Bj\"orken variable $x_A \equiv
Q^2/(2M_A\nu)$\cite{aku}. Generally the PDF and SF depends also on the resolution
$Q^2$ which is particularly important for $x_A<0.01/A$ where a nuclear shadowing takes
place. Shadowing should be included in any treatment of the EMC effect. However the
shadowing is described\cite{shadow,Weise,Kulagin} as a multi-scattering process with
diffraction between different nucleons. If the Momentum Sum Rule (MSR) has to be
analyzed,  the simple convolution of nucleon PDF with nuclear distribution preserves
the Longitudinal Momentum (LM) of this partonic system. For $x_A>0.1/A$ we
know\cite{Jaffe,shadow,Fran,Kulagin} that nuclear shadowing is unimportant.

In the Light Cone (LC) formulation\cite{Jaffe,Fran}, $x_A$ corresponds to the nuclear
fraction of a quark LM $k^+=k^0+k^3$ and is equal (in the nuclear rest frame) to the
ratio $x_A=k^+/P_A^+\equiv\sqrt{2}k^+/M_A$ - Lorentz invariant. But the composite nucleus
is made of hadrons which are distributed with longitudinal momenta $p_h^+$, where
$h=N,\pi,...$ stands for nucleons, virtual pions, ... . In the convolution
model\cite{Jaffe,Fran} a fraction of parton LM $x_A$ in the nucleus is given as the
product $x_A=x_h*y_h/A$ of fractions: a parton LM in hadrons $x_h\equiv Q^2/(2M_h\nu)=
k^+/p_h^+$ and a hadron LM in the nucleus $y_h=p_h^+/P_A^+$. The nuclear dynamics of
given hadrons in the nucleus is described by the distribution function $f_h(y_h)$ and PDF
$F^h_2(x\equiv x_h)$ describes its partonic structure. {Remember that there are two
different scales of interactions: long range nuclear scale which forms hadron
distribution functions in nuclear matter and a much shorter partonic scale which is
responsible for their PDF's.}
\subsection{Kinematics in the Bj\"{o}rken limit}
Consider  DIS on hadrons, $eH\rightarrow e'X$\cite{Jaffe,It}, as an introduction. In the
final state we measure the electron energy and scattering angle $\Theta$ of outgoing
electron. The virtual photon momentum transfer is:

\begin{equation}\label{q}
  q=(\nu,0,0,-\sqrt{\nu^{2}+Q^{2}}).
\end{equation}

\begin{figure}[t]
\includegraphics[width=7cm]{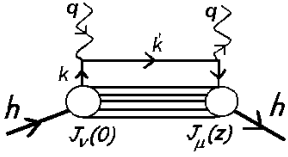}
\vspace{-25mm} \caption{~Hadron tensor $W\mu\nu$ in the parton model} \label{fig:Wmn}
\end{figure}

\noindent The differential cross-section
\begin{eqnarray}
d\sigma \sim L^{\mu\nu} W_{\mu\nu},
\end{eqnarray}
for electron scattering, in which hadrons in X states are not observed, is
proportional to the contraction of a lepton tensor $L^{\mu\nu}$ with the hadron tensor
$W_{\mu\nu}$ given by:
\begin{eqnarray}
W_{\mu \nu}\equiv \sum_{X}{(2\!\pi)}^4\delta^{4}\!(p\!+\!q\!-\!p_x)
<\!p|J_\mu(0)|X\!>\!<\!X|J_\nu(0)|p\!> \nonumber
\end{eqnarray}
\noindent where $J_\nu$ is a hadronic electromagnetic current operator. For low $Q^2$ one
could expect the corrections from strong interaction but for our nuclear purpose this
approximation is sufficient. Shifting the current $J_\mu(0)$ to the space-time point $z$
(see Fig.\ref{fig:Wmn}) and assuming completeness of intermediate states $X$ we get:
\begin{equation}
W_{\mu \nu} = \int d^4 \xi e^{iq \xi} <p|[J_\mu(\xi)J_\nu(0)]|0>. \label{W2}
\end{equation}
In the Bj\"orken limit, $Q^{2}=-q^2\rightarrow\infty$ and $q^{2}/\nu^{2}\rightarrow
0$, the scaling variable $x=Q^{2}/2M_h\nu\simeq k^{+}/p^{+}$ is fixed. For LC
components of $q$: $q^{-}=(q^{0}-q^{3})/\sqrt{2}\rightarrow\infty$ but
$q^{+}=-M_hx/\sqrt{2}$ remains finite. These imply for a conjugate variable $\xi$ in
Eq.\ref{W2}: $\xi^{+}\rightarrow0$ and $\xi^{-}\leq\sqrt{2}/M_hx$ from which one gets
the following restrictions for components:
\begin{eqnarray}\label{eq:zt}
  \xi_{0}\leq1/M_hx ~~~~~~~~~~ \xi_{z}\equiv z\leq1/M_hx.
\end{eqnarray}

\noindent The spatial variable $\xi$ is connected directly to the correlation length
in elementary the subprocess where the electron interacts with a quark and changes its
four momentum by \emph{q}. We have therefore two resolutions scales in a deep
inelastic scattering: $1/\sqrt{Q^{2}}$ which is connected with the virtuality of a
foton probe and $z=1/M_hx$ which is the distance how far the intermediate quark can
propagate in the medium, see Fig.\ref{fig:Wmn} and Eq.\ref{eq:zt}. Small x means a
relatively large correlation length $z$. The hadron tensor $W_{\mu \nu}$ can be
expressed  in terms of two structure functions $W_1$ and $W_2$ depending of two
Lorentz invariants: $q^2$ and \emph{pq=}$M_h\nu$. In the Bj\"orken scaling, the DIS is
described\cite{It} by the PDF - $F_2(x)$ defined in a target rest frame\cite{Jaffe},
where $p^+\!=\!M_h/\sqrt{2}$, in terms of quark $q_{n_f}(x)$ and antiquark
$\overline{q}_{n_f}(x)$ distributions:

\begin{eqnarray}
F_2(x)\!=\!\frac{\nu}{M_h}~\!\lim_{_{Bj}}
W_2(q^2,\nu)\!=x\!\sum_{n_f}(q_{n_f}(x)\!+\!\overline{q}_{n_f}(x))
\end{eqnarray}
They are given  by LC fields $\psi_+$ with the sum over flavors $n_f$:
\begin{eqnarray}
\begin{array}{cc}{q_{n_f}}(x)\\\overline{q}_{n_f}(x)\end{array}
 =\sum_{X} \delta(p^+\!-\!xp^+\!-\!p_X^+) {\left| <X|
 \begin{array}{cc}\psi_{+n_f}\\\psi_{+n_f}^\dag\end{array}
| p> \right|}^2 \nonumber \label{finalform}
\end{eqnarray}
. Functions $q_{n_f}(\overline{q}_{n_f})$ are probabilities to remove the quark
(antiquark) with flavor $n_f$ from target living behind some remnant with the momentum
$(1-x)p^+$. Resulting sum rule for a total number of quarks $N_q$ in a hadron and MSR
for its total LM fraction $M_q$ are:
\begin{eqnarray}
\int_0^1\! F_{2}(x)\frac{dx}{x}=N^0_q~~~\mbox{and}~~~\int_0^1\!F_{2}(x)dx=M_q.
\label{constituents}
\end{eqnarray}
Using an additional phenomenological observation that in hadrons and in good
approximation in nuclei (EMC effect) the LM is equally distributed between quarks and
gluons we can normalize $M_q\!=\!1$ in order to get the total LM. Such a MSR should be
satisfied because only partons, constituents of the strongly interacting system, carry
the whole LM of a nucleon or a nucleus.
\subsection{A Convolution Model in the Nuclear Mean Field}
 Each parton carries its $x_A$ fraction of LM
and the nuclear SF $F^{A}_{2}(x_A)$ which described  the distribution of this fractions
is normalized. The Bj\"orken scaling in $x_A$ corresponds in the LC dynamics\cite{Jaffe}
to the scaling in $x$ for a  nucleon PDF  $F_2^N(x)$. This scaling is caused by the
relativistic contraction of nucleons. In the convolution model restricted to nucleons and
pions (lightest virtual mesons) the nuclear SF $F^{A}_{2}$ is described by:
\vspace*{-2mm}
\begin{eqnarray}
\!\!\!F^{A}_{2}\!(x_A)\!\!&=&\!\!\!\sum_{h=N,\pi}\!\int\!{ydy}\!\!
\int\!\!{\!dx_h}\delta(x_A\!-\!x_hy_h)f_h\!(y)F^{h}_{2}\!(x)
\label{structure}\\
f_N(y)\!\!&=&\!\int\!\!{d^4p\over(2\pi)^4}\delta(y-{A(p^0+p^3)\over{M_A}}) Tr\!\left[
{\gamma^+}S(p,P_A)\right]\nonumber
\end{eqnarray}

\noindent where $y=Ay_N$ and $F_2^{N}$ will be later replaced by  $F_2^{B}$ the PDF of
bound nucleon. Both quark and nucleon distributions are manifestly covariant and can be
expressed by Green's functions\cite{Jaffe} $S(p,P_A)$. The trace is over the Dirac and
isospin indices and the single nucleon Green's function in the nuclear medium is given
i.e. in \cite{wa,ms2}:
\begin{eqnarray}
S(p,P_A)=-i\left(\gamma\cdot(p-U_V)+M_N^* \right)\times\label{gf}
\end{eqnarray}
\begin{eqnarray}
\left[\frac{1}{(p-U_V)^2\!-\!{M_N^*}^2\!+\!i\epsilon } +
\frac{i\pi\theta(p_F\!-\!|\textbf{p}|) }{E^*_N( p)} \delta (p^0\!-\!E^*_N(p)\!-
\!g_VU_V^0)\right]  \nonumber
\end{eqnarray}
where
\begin{equation}
E^*_N(p)\equiv\sqrt{{M_N^*}^2+{\bf p}^2}~~~~~~~~~~~~~M^*_N \equiv M_N+g_S U_S.
\nonumber
\end{equation}
The effective mass $M^*_N$ is substantially lower from the bare nucleon mass $M_N$
(vacuum value). The values of vector $U_V\!=\!g_V(U_V^0,\textbf{\emph{0}})$ and scalar
$g_SU_S$ potentials are discussed for example in four specific mean-field
models\cite{wa,ser,boguta,zm}.
The connected part (second term) of (\ref{gf}) inserted into Eq.\ref{structure} for
$f_N(y)$ gives, after taking the trace and using the delta function to integrate over
$p^0$, the result which can be simplified in the RMF to the form\cite{Mike}:
\begin{eqnarray}
f_N(y)\!&\!=\!&\!\frac{4}{\varrho}\int_{|p|<p_F}\frac{S_N(p)d^3p}{(2\pi)^3}
(1\!+\!\frac{p_3}{E^*_N})\,\delta(y\!-\!p^+/\varepsilon_A) \nonumber
\\&=& \frac{3}{4}\left(\frac{\varepsilon_A}{p_F}\right)^3 \left[
\left(\frac{p_F}{\varepsilon_A} \right)^2-\left(y-\frac{E^A_F}{\varepsilon_A}
\right)^2 \right], \label{RMF}
\end{eqnarray}
Here the nucleon spectral function was taken in the impulse approximation:
$S_N=n(p)\delta (p^o-(E_N^{^{*}}(p) +U_V^0))$, $\varepsilon_A=M_A/A$. $E_F^A$ is the
nucleon Fermi energy and $y$ takes the values determined by the inequality
$(E^A_F-p_{F})/\varepsilon_A<y<(E^A_F+p_{F})/\varepsilon_A$. Finally the nucleon
distribution function $f(y)$ depends on its Fermi momentum $p_{F}$, Fermi energy
$E_F^A$ and a single particle energy $\varepsilon_A$ but only two of them or
independent in RMF.
 Using Eqs.\ref{constituents},\ref{structure},\ref{RMF} with $F^{B}_{2}\!=\!F^{N}_{2}$ and neglecting pion contributions we
obtain:
\begin{equation}
\int\!dx_A\,F_2^{A}(x_A)\! =\!\int\!dy\,yf_N(y)=\frac{E^A_F}{\varepsilon_A}\geq1
\label{RMF2}
\end{equation}
where the last inequality obtained for $p_H\geq0$ comes from the following HvH relation
between $E_F^A$, $\varepsilon_A$ and NM pressure $p_H$ (see for example \cite{kumar})
which was proven in the self consistent RMF approach \cite{boguta}. According to HvH
theorem a Fermi energy given as baryon density derivative of a energy
$E^A=A\varepsilon_A$ in a constant volume $\Omega$ is:
\begin{eqnarray}
\hspace{-5mm} E^A_F\equiv\frac{\partial}{\partial
A}\left(E^A\right)_{\Omega}=\frac{d}{d\varrho} \!\left(\!\frac{E^A}{\Omega}\!\right)
\label{ef} = \varepsilon_A\!+\!\frac{p_H}{\varrho}=\varepsilon_A\!+\!E_{press}
\nonumber \label{RMF3} \hspace{-10mm}
\\
\end{eqnarray}
where $\varrho=A/\Omega$ and the hadron pressure $p_H$ is given by the thermodynamic
relation:
\begin{eqnarray}
p_H&=&-\left( \frac{\partial E^A}{\partial \Omega}
\right)_{\!\!A}\!=\varrho^2\frac{d}{d\varrho}\left(\varepsilon_A \right).
\label{ther}
\end{eqnarray}
The integral (\ref{RMF2}) is equal to $1$ at the saturation point. Taking only a
nucleon contribution in (\ref{structure}) it would mean that nucleons carry at
equilibrium the whole $P_A^+$ of the nucleus although mesonic fields $(U_S, U_V)$ are
very strong (few hundred MeV)\cite{ms2}. But we know\cite{Mike} that pure Fermi motion
can not describe the EMC effect therefore $F^{B}_{2}(x)\neq F^{N}_{2}(x)$. When the
resolution $z\!=\!1/(M_Nx)$ (Eq.\ref{eq:zt}) is smaller then one half of the  NN
distance $d\!\simeq\!1/\varrho^{\frac{1}{3}}$, the single particle area
 is "visible" in DIS\cite{jacek}. Let us determine the
limiting value $x_{L}$ of $x$ from $z$:
\begin{equation}
 x_{L}=2/(M_N d)\simeq2\varrho^{\frac{1}{3}}/M_N \sim0.25.
\end{equation}
The EMC ratio $(\sigma_A/\sigma_D\!\simeq F^{A}_{2}\!(x)/F^{N}_{2}\!(x))$ shows
\cite{newdata} for $0.3 \lesssim x \lesssim 0.6$, that $F^{A}_{2}$ and consequently
$F^{B}_{2}$, is gradually smaller from a free PDF $F^{N}_{2}$.
 The average ($p_3=0$) LM fraction $x_{max}$
taken by partons localized in nuclear pions should not exceed the mass ratio
$x_{max}\approx M_\pi/M_N\simeq0.15$. Clearly for small $x\leq0.25$ where we expect
contributions from pionic partons we can observe a small excess in EMC
ratio\cite{newdata}. If MSR is satisfied, the excess for small $x$ can be interpreted
as contributions of nuclear pions carrying LM while being exchanged between separate
nucleons seeing in the large $x\gtrsim x_L$ region mentioned before. The resulting
average pion excess $<\!{p_{\pi}}^{+}\!>$ included formally in (\ref{structure}) is
given by the difference:
\begin{eqnarray}
\frac{<p_{\pi}^+\!>}{<\!P_A^+\!>}\!\!&=&\!\!\!\int^1_{x_L}\!\left(F^{B}_{2}(x)-\!F^{N}_{2}(x)\!\right)\!dx
 \label{sum}
\end{eqnarray}
Phenomenologically this ratio is sufficiently small \cite{drell,newdata,ms2} ($\sim 1\%$)
to describe also the nuclear Drell-Yan reactions. So finally the nuclear MSR:
\begin{equation}
\int\!dx_A\,F_2^{A}(x_A)\! =1 \label{MSR}
\end{equation}
is satisfied\cite{RW} for $p_H=0$ with a modified PDF $F^{B}_{2}\!\neq\!F^{N}_{2}$ and a
small admixture of virtual pions. For $x\!\simeq\!0.5$, where eventually the excess of
heavier meson would be present there is a clear reduction of a nuclear SF. (The Fermi
motion starts to increase EMC ratio for $x\geq0.7$.)
\begin{figure}
\hspace{-4mm}
\includegraphics[height=7cm,width=8.9cm]{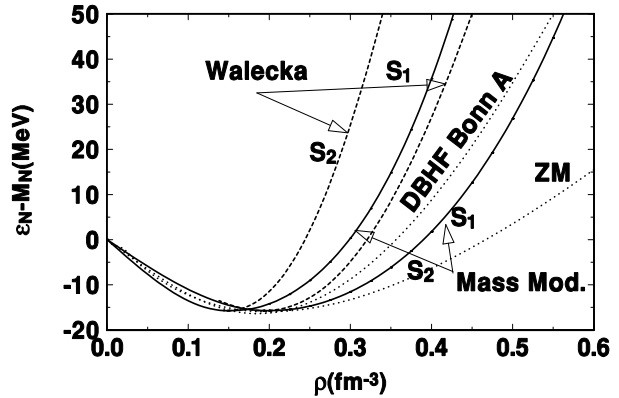}
\vspace{-1mm} \caption{The nucleon energy $\varepsilon_A-M_N$ as a function of NM density
for two RMF models; $\sigma-\omega$ Walecka (dot lines) and our Modified Mass approach
(solid). Both RMF models are calculated for two parameterizations: $S_1$
version\cite{ser} $(\varrho_0=.16fm^{-3})$ and version $S_2$\cite{wa}
$(\varrho_0=.19fm^{-3})$. Results for full DBHF\cite{bonn} (dotted marked line)
calculation using Bonn A NN interaction are displayed for comparison, also nucleon energy
in ZM model\cite{zm} is in the plot (dotted marked line).} \label{fig:enernew}
\end{figure}
\section{Non-equilibrium corrections to nuclear distribution} A nucleon repulsion and increasing
pressure distorts the parton distribution in nucleons. For example, describing
nucleons as bags, the finite pressure will influence their surfaces
\cite{Koch,Hua,bag,Kap,Jennings}.
In the paper we show how such a modification of PDF will influence the EoS.

Consider the nuclear pion contributions above the saturation point.  In
Dirac-Brueckner calculations the pion effective cross section in a reaction
$N+N=N+N+\pi$ is strongly reduced at higher nuclear densities above the threshold
\cite{hm1} ( also with RPA insertions to self energy of $N$ and $\Delta$ \cite{oset}
included). Moreover the average distances between nucleons are smaller and nucleons
approach eventually close packing limit. In fact the limiting parameter $x_{L}$ in
(\ref{sum}) will increase with a density and the room for nuclear pions given by
(\ref{sum}) will be reduced. Summarizing, the separated nuclear pions carry possibly
less then $1\%$ of the nuclear LM for positive pressure and dealing with a
non-equilibrium correction to the nuclear distribution we will restrict considerations
to the nucleon part (h=N in Eq.\ref{structure}) without additional virtual pions.
The eventual admixture of additional pions makes the violation of longitudinal momentum
even stronger.

The Equation of State (EOS) for NM has to match the saturation point with
compressibility $K^{-1}=9\varrho^2\frac{d^2}{d\varrho^2} \frac{E}{A}$ for
$\varrho=\varrho_0$ but then the behavior for higher densities is different for
different RMF models. Generally, the choice of initial Lagrangian in nuclear RMF
models and the dependence of nucleon masses from a density is not unique. Let us
compare two density dependent effective masses from extremely different examples of
RMF models. It is well known that in the linear W model the compressibility defined at
the saturation density is too large $K^{-1}\simeq560$ MeV. The non-linear
Zimanyi-Moszkowski (ZM) model produces the soft EoS with a good value of
$K^{-1}=225MeV$.  In both models, two coupling constants of the theory are fixed at
the semi-empirical saturation density of NM. In the stiff W model
$M^*_N\!\simeq\!0.6M_N$ at equilibrium. Here effective mass $M^*_N$ is obtained by a
subtraction of strong scalar field from the nucleon mass at a saturation point (for a
respective EOS see marked lines in fig.\ref{fig:enernew}). In a model
\cite{zm,DCMP,GWM} the fermion wave function is re-scale and interprets the new,
density dependent effective "Dirac" mass $M_{ZM}$. It also starts to decrease with a
density from $\varrho=0$, and at the saturation point reaches $~85\%$ of a free
nucleon mass (there is also extended ($\sigma-\omega$) model which include self
interactions of the $\sigma$-field\cite{stocker} with 2 additional parameters).
 The nucleon mass $M_N$ replaced at the saturation point by
smaller masses $M^*_N$ or $M_{ZM}$ would change significantly the PDF, shifting the
Bj\"orken $x\sim(1/M_N)$. From the point of view of the description of the EMC effect
\cite{aku,ms2,jacek,RW}, this means that nucleons will carry $(15-40)\%$ less of a
 LM,  what should be compensated
by the enhanced contribution from a meson cloud for smaller $x$. There is no evidence
for a such huge enhancement in the EMC effect for small $x$. This complicates the
simple description of the EMC effect\cite{ms2,newdata} with the observation\cite{ms2}
of some effect involving dynamics beyond the conventional nucleon-meson treatment of
nuclear physics. The "EMC" effect certifies that a departure from the free PDF is
rather small (only few percent), also MSR (\ref{MSR}) is satisfied within $1$\%. The
nuclear Drell-Yan experiments\cite{drell,ms2} which measure the sea quark enhancement
are described simultaneously\cite{jacek} with such a small $1$\% admixture of nuclear
pions and the nucleon mass unchanged.
\begin{figure}[t]
\vspace{-6mm}
\begin{center}
\includegraphics[height=6.5cm,width=8.cm]{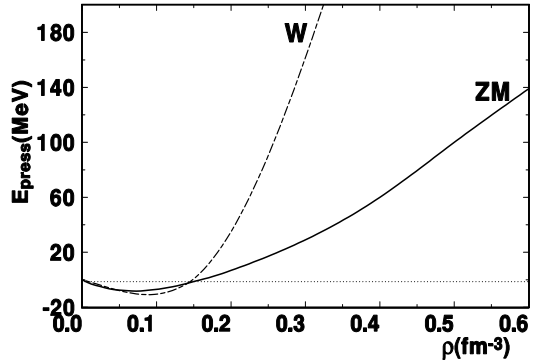}
\end{center}
\caption{Corrections $E_{press}=\frac{p_H}{\varrho}$ in the evolution of the PDF
inside NM for stiff W($S_1$) and soft ZM models (comp. Fig.1).} \label{meson}
\end{figure}

Introduce a nucleon mass in medium $M_{med}$ defined as the total LM of partons in a
nucleon rest frame. If  at the equilibrium weekly bound nucleons carry almost a whole
average $<\!P_A^+\!>$ then the nucleon mass in medium should be close $M_{med}\simeq M_N$
to its vacuum value. Above the saturation point the increasing pressure between nucleons
starts to increase the
 $E_F^A/\varepsilon_A$ (\ref{RMF3}) thus MSR (\ref{MSR}) is broken by $E_{press}/\varepsilon_A$ (\ref{RMF2}).
 $E_{press}$ calculated for those two models is shown on
 Fig.\ref{meson}. It is relatively small (therefore the violation of MSR is weaker) in ZM model (for
$\varrho=0.3fm^{-3}$ only $25$ MeV in comparison to $150$ MeV in W model) but not
negligible. This "unexpected" strong departure of the MSR (\ref{RMF2}) from $1$ for a
positive pressure $p_H$ in the NM originates from a relativistic flux
factor\footnote{In a criticized\cite{Fran} non-relativistic approach always
 $\int\!dy\,yf_N(y)\!=\!1.$} $(1\!+p_3/E^*_N)$ included in
 (\ref{RMF}) and not from a version of the RMF model. Consider modifications of the function $F^{B}_{2}\!(x)$
 along with a pressure to  fulfill
 the condition (\ref{MSR}).
\subsection{The nucleon PDF for finite pressure}
In the RMF the nucleons are approximated by point like objects, which interact
exchanging mesons. But in fact nucleons have a finite volume therefore a positive
pressure should influence internal parton distributions. This process can not be
described clearly by a perturbative QCD[30] but a next subsection contains a simple
bag model estimate. Partons - gluons and quarks inside compressed nucleon will start
to adjust their momenta to nucleon properties like a surface, volume and a mass. The
particularly energetic weakly bound partons give a large contribution to the nucleon
rest energy. Simultaneously, partons take part in the increasing Fermi motion of
nucleons.  These squeezed extended objects exist in NM  under a positive pressure and
the amount of energy is required to make a room for nucleons by displacing its
environment. It will reduce the sum of $N_q$ parton LM momenta in medium:
\begin{equation}
\sqrt{2}\!\!\!\!\sum_{medium}^{N_q}\!\!\!\!\!k_{i}^+=\!\!\!\!\sum_{medium}^{N_q}\!\!\!\!\!k_{i}^0
<M_N\!=\sqrt{2}p^+~~~~\mbox{or}~~\int_0^1\!\!\!dx\,F_2^B(x)<1. \label{dupa1}
\end{equation}

\noindent Let us compare two inequalities  (\ref{RMF2},\ref{dupa1}) induced by a
pressure, with the total MSR (\ref{MSR}). We propose to use  (\ref{dupa1}) the
inequality for the parton LM distribution in a nucleon in medium in order to meet the
total LM momentum sum rule (\ref{MSR}) violated linearly (\ref{RMF2},\ref{RMF3}) along
with a positive pressure. To this end the inequality (\ref{dupa1}) with the PDF
$F_2^B(x)$ has to fulfill the condition:
\begin{equation}
\int_0^{1}\!dx\,F_2^B(x)=\frac{\varepsilon_A}{E^A_F}\leq1~~\mbox{for}~~p_H \geq 0
\label{dupa2}
\end{equation}
in order to satisfy, with the help of (\ref{RMF2}), the MSR:
\begin{eqnarray}
\!\int^1_0\!F^{A}_{2}(x_A)dx_A\!=\!\frac{E^A_F}{\varepsilon_A}\!\int_0^1\!
F^{B}_{2}(x)dx=1\label{MSR1}
\end{eqnarray}
where $x\!=\!{k^+}/{p^+}$. To estimate main effects in medium, we assume the following
parametric form of a PDF in NM: $F^{B}_{2}\!(x)=aF^{N}_{2}\!(bx\!)$. The required
condition (\ref{dupa2}) determines the relation $b=a E^A_F/\varepsilon_A$ therefore we
have:
\begin{eqnarray}
F^{B}_{2}\!(x)=aF^{N}_{2}\!\left(a\frac{E^A_F}{\varepsilon_A}x\!\right) \label{scale}
\end{eqnarray}
with a free parameter $a(p_H\!)$. The number of valence quarks should not be changed.
However a total parton number (\ref{constituents},\ref{dupa2}) $N_q=a(p_H)N^0_q$ will
eventually increase in a more energetic compressed medium, thus $a\!\geq\!1$; e.q. a
simple choice $F^{B}_{2}\!(x)\!=\!({\varepsilon_A}/{E^A_F})F^{N}_{2}\!(\!x\!)$ is not
suitable; although satisfies Eq.(\ref{dupa2}), it decreases
 $N_q$. The number of quark constituents $N_q$ (which
include sea quarks) is preserved for $a\!=\!1$. The scaling of Bj\"orken $x$ by the
factor $a{E_F^A}/{\varepsilon_A}$ squeezes a PDF in NM towards smaller $x$,
consequently the sum of the quark longitudinal momenta given by the PDF integral
(\ref{dupa2}) is smaller. The actual upper limit in (\ref{dupa2}) is diminished (see
(\ref{scale})) to $x_{up}=\varepsilon_A/(aE^A_F)$. Thus $F^{B}_{2}(x)$ is assumed to
be negligible for large $1>x>x_{up}$.

In a RMF approach the detailed form of the nucleon PDF ({\ref{scale}) with the
specified parameter "$a$" or $N_q(p_H)$ is not important for EOS; important is the
condition ({\ref{dupa2}) which defines in the nucleon rest frame a nucleon mass in
medium $M_{med}$ with the following decrease along with the increasing pressure $p_H$:
\begin{eqnarray}
M_{med}\!&\equiv&\!\sqrt{2}\!\!\sum_{medium}^{N_q}\!\!\!\!\!k_{i}^+ =M_N\!\!\int\!dx
F^{B}_{2}\!(x)\label{masseff} =\!M_N\frac{\varepsilon_A}{E^A_F}\\&=&M_N\!/\!
\left(\!1\!+\!\frac{p_H}{\varrho} \right)\!\simeq\!M_N\!\left(
\!1\!-\!\frac{p_H}{\varrho\varepsilon_A}\right)~~\mbox{for}~~p_H>0\nonumber
\end{eqnarray}

\noindent Concluding, changes  of the nucleon PDF (\ref{dupa2}) affect the nucleon
mass in a medium (\ref{masseff}) setting $M_{med}\leq M_N$.

Please note that for $\varrho<\varrho_0$ nucleons are well separated, therefore we assume
that the nucleon PDF and mass remain unchanged. However for $p_H<0$ the MSR integral
(\ref{RMF2}) $(1\!+\!E_{press}/\varepsilon_A)\!<\!1$ (see Fig.\ref{meson}). The missing
negative part: ($E_{press}/\varepsilon_A$) of  LM  is taken in our approach by nuclear
pions (\ref{structure}). Its biggest contribution (\!$\sim1\!\%$) is obtained (depending
from the RMF model) for $\varrho\!\simeq\!(0.05\!-\!0.1)$ and disappears along with the
$p_H$ for $\varrho\!\rightarrow\!0$.

\subsection{The bag model estimate }
Let us discuss these mass modifications in the simple bag model\cite{MIT} where the
nucleon in the lowest state of three quarks is a sphere of volume $\Omega_{N}$ and its
energy $E_{Bag}$ is given in a vacuum as a function of a radius $R$ with
phenomenological constants - $\omega_0$, $Z_0$ and $B$: \vspace{-2mm}
\begin{eqnarray}
E^0_{Bag}\!(R)\!\!&=&\frac{3\omega_0-Z_0}{R}+\frac{4\pi}{3}BR^3\sim~1/R \label{bag}
\end{eqnarray}
 The following condition for the
pressure $p_B=0$ inside a bag in equilibrium gives the relation between $R$ and $B$
which was used in
 the last relation of (\ref{bag}):
\begin{eqnarray}
p_B=\left(\partial E_{Bag}/\partial \Omega_{N}\right)_n =0 \label{pressured2}
\end{eqnarray}
$E^0_{Bag}$ differs from the nucleon mass by the c.m. correction
\cite{Jennings} in the partonic model of a nucleon.

However in a compressed medium the pressure generated by free quarks inside the bag is
balanced at the bag surface\cite{MIT} not only by a intrinsic confining force
represented by the bag "constant" $B(\varrho)$ (which depends on $\varrho$) but
additionally by a NM pressure $p_H$ generated by elastic collisions with other
hadron\cite{Koch,Kap} bags or a NN pressure derived in QMC/QHD model in
medium\cite{bag}. Using a spherical bag solution (the first relation in (\ref{bag}))
with the formula (\ref{pressured2}) for finite $p_B$ we can obtain the expression for
the radius $R$ of a compressed nucleon. Now the pressure $p_B$ inside a bag is equal
on the bag surface to an external pressure $p_H$ and finally:
\begin{eqnarray}
p_H\!=p_B\!\!&=&\!\!\frac{3\omega_0-Z_0}{4\pi
R^4}-\!B(\varrho)~~\rightarrow~~(B(\varrho)\!+\!p_H\!)R^4\!=\!const \nonumber
\label{pressure}\\
R\!&=&\!\left[\frac{3\omega_0-Z_0}{4\pi
(B(\varrho)+p_H)}\right]^{1/4}\label{rsolution}
\end{eqnarray}
The pressure $p_H$ between hadrons acts on the bag surface similarly to the bag
constant $B$. At the saturation $p_H=0$ and the bag "constant" $B(\varrho_0)$ is
determined by the value of the nucleon radius $R\simeq1fm$. Above the saturation point
when the NM pressure $p_H$ would be not taken into account ($p_H=0$ in
(\ref{rsolution})) the nucleon radius $R$ increases\cite{Jennings} in a NM. It is
shown\cite{bag} that an decreasing of the B constant from a saturation density
$\varrho$ up to $3\varrho$ by $~60MeVfm^{-3}$ is accompanied be similar increase of
the pressure $p_H$. The changes in medium depend on the EoS.  The QMC model in
medium\cite{Hua} takes into account the $p_H$ contributions to the bag radius. In
particular for the ZM model which has the realistic value of $K^{-1}=~225MeV$ the
nucleon radius remains almost constant\cite{Hua} up to density $\varrho=10\varrho_0$.
Such a solution of a slowly varied $R$ with $\varrho$ is probably the property of the
relatively soft EoS\cite{Hua}; e.q. a $ZM$ model shown in Fig.\ref{fig:enernew}. Also
in our estimate (\ref{rsolution}), when the bag radius weakly depends from the
increasing density the sum $(B(\varrho)+p_H)$ remains approximately constant.

The nucleon rest energy $E_{Bag}$ under the compression $p_H$ can be finally obtain from
(\ref{bag},\ref{pressured2},\ref{rsolution}):
\begin{eqnarray}
E_{Bag}\!\!\!&=&\!\!4\pi\! R^3\!\left[\frac{4}{3}(B+p_H)\!-\!\frac{p_H}{3}\right]\!=
\!E_{Bag}^{0}\frac{R_0}{R}\!-\!p_H\Omega_{N} \nonumber \\
\label{massbag}
\end{eqnarray}
where $R_0$ and $E_{Bag}^{0}$ denote a radius and a bag energy fit to the nucleon mass
for $p_H=0$. The scaling factor $R_0/R$ comes from the well known model dependence
(\ref{bag}) ($E_{bag}^0\!\sim\!1/R$) in a spherical bag \cite{MIT}. This simple radial
dependence is now lost in (\ref{massbag}).

Responsible for that is the pressure dependent correction to a mass of the nucleon
given by the product of $p_H$ and the nucleon volume $\Omega_N$. Now we can compare it
with a similar correction to the nucleon mass from (\ref{masseff}). They have a common
linear behavior with the pressure and they are equal for
$\varrho\!\simeq\!(M_N/\varepsilon_A)\varrho_{max}$ - where
$\varrho_{max}=1/\Omega_{N}$ denotes the greatest density of not overlapping
(approximately) nucleon bags. These corrections are express above all by a product of
pressure and a single particle volume $\Omega_N$ (\ref{massbag}) or $\Omega/A$
(\ref{masseff}) which physically means the necessary work W=$p_H\Omega_N$ to be done
in order to create a space for this extended system - the nucleon in a compressed NM.
These RMF results shows that the nucleon mass $M_N$ can be consider generally as the
enthalpy $H=U+p_H\Omega_N$ - equal to the total energy which includes the $M_{med}$
(as an internal energy $U$) and a work $W$. Thus $M_N\simeq M_{med}+p_H\Omega_N$. In
the nuclear medium in equilibrium $p_H\!=\!0$ therefore $M_{med}\!=\!M_N$.
\begin{figure}\includegraphics[height=6cm,width=8.cm]{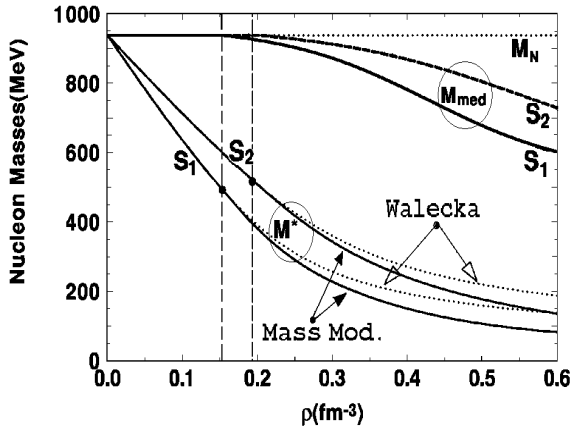}
\caption{The constant nucleon mass $M_N$ used in W model and the density dependent
mass $M_{med}$ from our "Mass Mod." model. Also respective effective mass $M_N^*$ and
$M_{med}^*$. Both models are calculated for $S_1$ and $S_2$ parametrization. Vertical
lines indicate saturation points.\label{fig:mass}}
\end{figure}
\vspace{-5mm}
\section{Results}
Our calculations show how changes in nucleon mass, will soften the stiff EoS of linear W
model\cite{wa} shown on Fig.\ref{fig:enernew}. In our calculations we replace the nucleon
mass $M_N$ by the mass in medium $M_{med}$, see Fig.\ref{fig:mass}. To accomplish it, our
explicit mass dependence (\ref{masseff}) from density, energy $\varepsilon_A$ and
pressure $p_H$ is combined with the standard linear RMF equations \cite{wa} for the
energy per nucleon $\varepsilon_A$ in terms of effective mass $M_{med}^*$(analogous to
$M^*_N$ in Eq.\ref{gf}):
\begin{eqnarray}
\varepsilon_A\!\!=\!C_1^2\varrho
\!\!\!&+&\!\!\!\frac{C_2^2}{\varrho}\!(M_{med}\!-\!M_{med}^*)^2\!\!+
\!\frac{\gamma}{\varrho}\!\!\int_0^{p_F}\!\!\!\frac{d^3p}{(2\pi)^3}\sqrt{p^2\!+\!{M_{med}^{*2}}}
\nonumber\\
M^*_{med}\!\!\!&=&\!\!M_{med}\!-\!\frac{\gamma}{2C_2^2}\int_0^{p_F}\!\!\!\frac{d^3p}{(2\pi)^3}
\frac{M^*_{med}}{\sqrt{p^2\!+\!M^{*2}_{med}}} \label{eq}
\end{eqnarray}
where $\gamma$ denotes a level degeneracy ($\gamma=2$ for a
neutron matter) and two (coupling) constants: vector $C_v^2$ and
scalar $C_s^2$,
 were fitted\cite{wa,ser} at the saturation point of nuclear matter (in the formula
$2C_1^2=C_v^2/M_N^2$, $2C_2^2=M_N^2/C_s^2$ with $g_VU^0_V=2C_1^2\varrho )$. \noindent
In the direct coupled $W$ model the nucleon mass $M_N$ is constant. In our modified
version the finite pressure corrections to $M_{med}$ (\ref{masseff}) convert the
recursive equation (\ref{eq}) to a differential-recursive set of equations above the
saturation density $\varrho_0$ in a general form:
\begin{equation}
f(\varepsilon_A,\frac{d}{d\varrho}\left({\varepsilon_A}\right))=0
~for ~\varrho\geq\varrho_0 \label{eq2}
\end{equation}
Note that equation (\ref{eq}) is obtained from the energy-momentum tensor for the model Hamiltonian with
a constant nucleon mass\cite{wa}. Here we assume that the same equation  with a medium mass $M_{med}$
will be satisfied. It should be a good approximation, at least not very far from the saturation density.
The pressure $p_H$ is obtained from the thermodynamic relation (\ref{ther}).
\begin{figure}[b]
\vspace{-4mm}
\includegraphics[height=6cm,width=8.6cm]{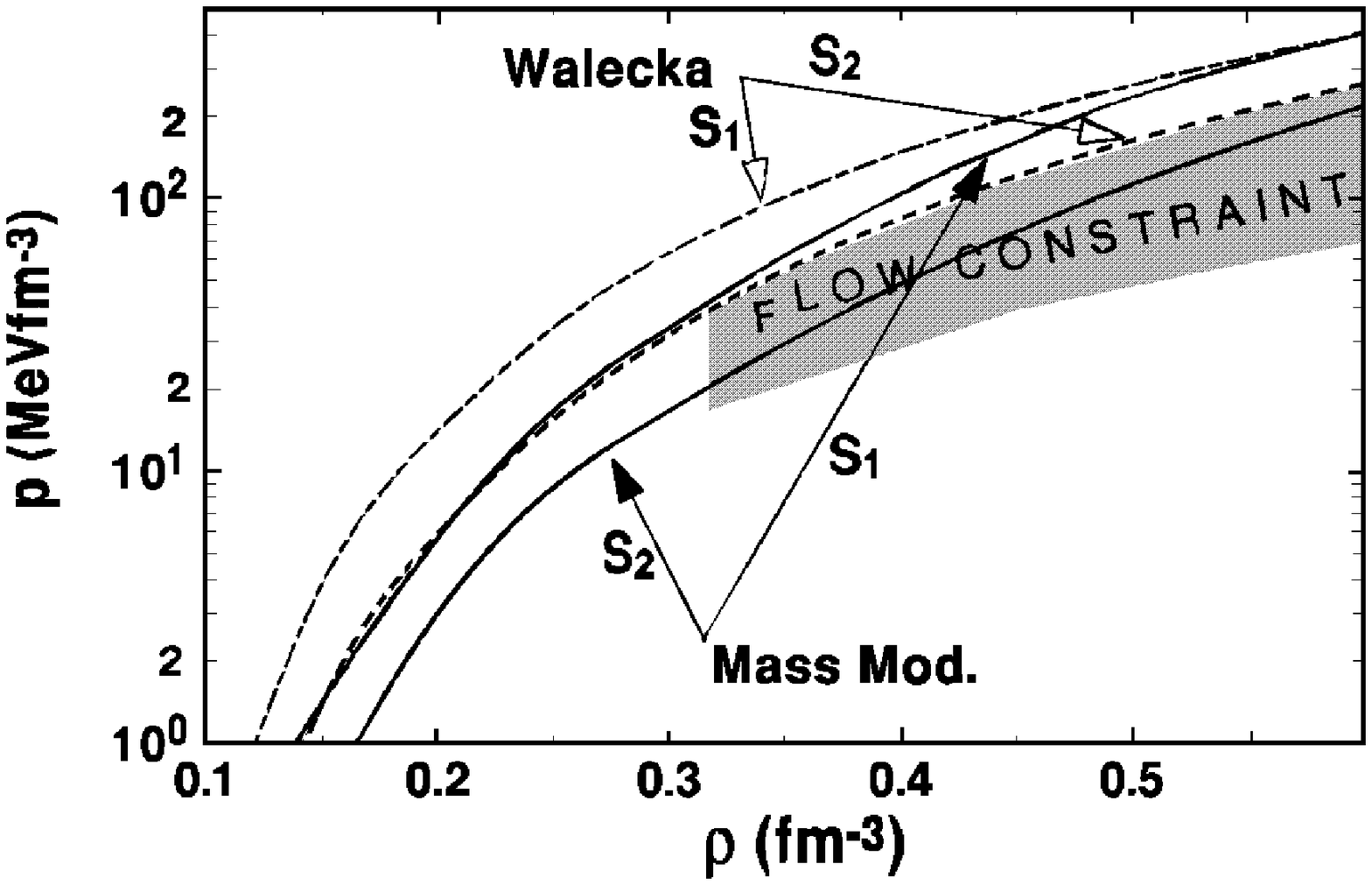}
\hspace*{-2mm} \vspace{-3mm}
\includegraphics[height=6cm,width=9.1cm]{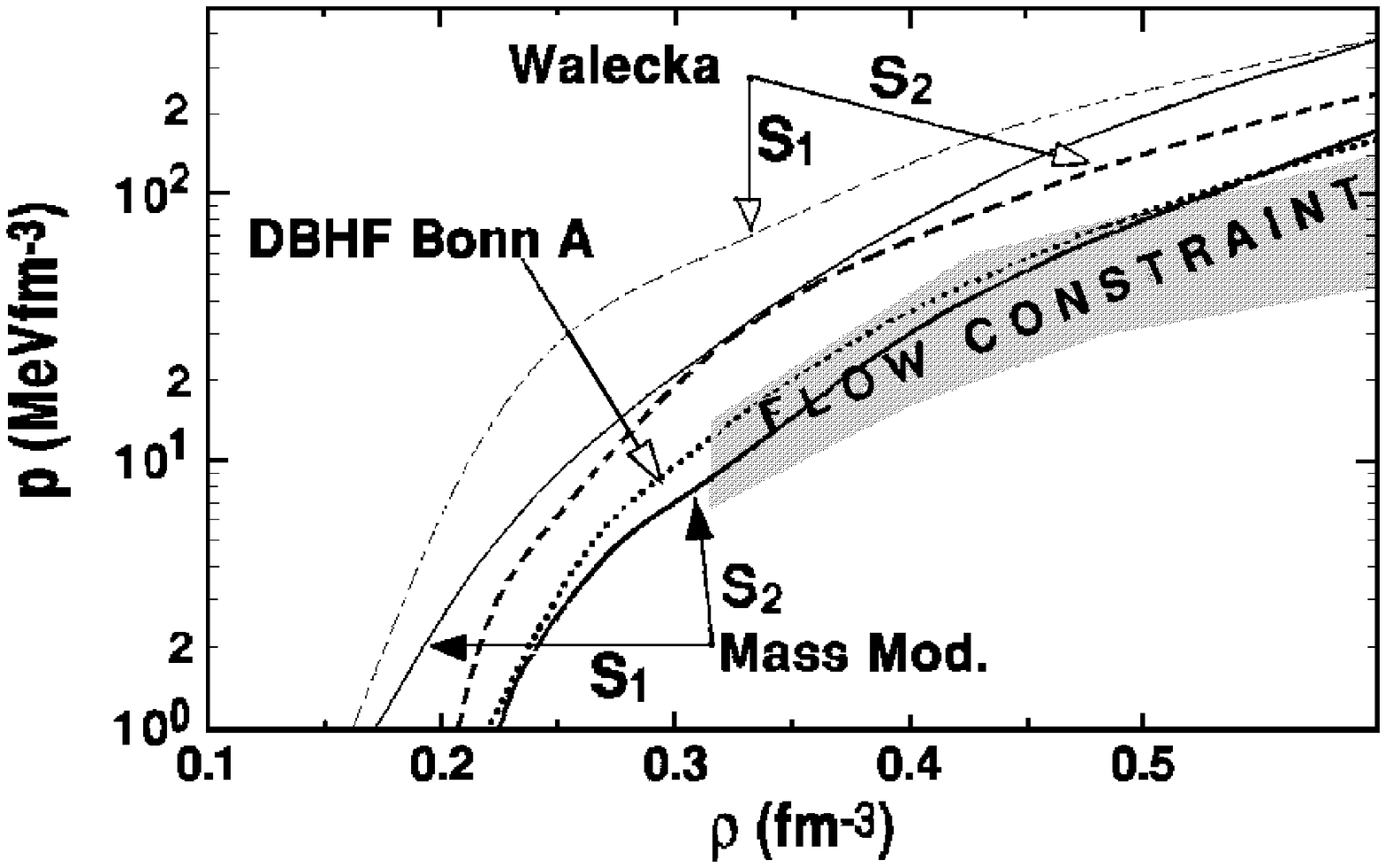}
\caption{The pressure for the neutron matter (upper plot) and for the nuclear matter (bottom)
 as a function of the density for two most frequent  parameterizations $S_1$ and $S_2$
 (see Fig.1 captions) of  W model is denoted by dashed lines. Our results for these parameterizations (Mass Mod.) are
 denoted by solid lines. The area denoted by "flow
constraint" taken from\protect\cite{pawel} determined the allowed course of EoS, using
the analysis which extracted from matter flow in heavy ion collisions the high
pressure obtained there. The DBHF (ref.\cite{bonn}) calculation with Bonn A
interaction are shown as a dotted line.} \label{main}
\end{figure}
The final results  were obtained
 by solving\footnote{It is
important to mention that in solutions of Eqs.(\ref{eq},\ref{eq2}) the Fermi energy from
definition Eq.(\ref{ef}) has a different value then the one calculated from the usual
form $E_F^A=\sqrt{{M^*_N}^2+p_F^2}+U_v$ used in Eq.(\ref{RMF}). The discrepancy vanish
near the saturation density, increases with the density and reach the $~15\%$ of the
total vector repulsion in Eq.(\ref{eq}). Similar problems\cite{NM} are connected with the
proper choice of single particle potential. which in our case should be adjusted to the
changes of nucleon mass. This discrepancy can be removed, here e.q. by the less repulsive
momentum dependent vector potential for nucleons however such a correction has no
influence on presented results.} numerically differential recursive equations
(\ref{eq2}), starting from standard solutions of Eq.(\ref{eq}) at the saturation density
for two version of the Walecka model: a first version $S_1$\cite{ser}
$(\varrho_0=.16fm^{-3}, {C_v}^2=273.8, {C_s}^2=357.4)$ and a second version
$S_2$\cite{wa} have a minimum at $\varrho_0=.19fm^{-3}$ (parameters ${C_v}^2=195.9,
{C_s}^2=267.1)$. They are displayed in
Figs.(\ref{fig:enernew},\ref{fig:mass},\ref{main}). In Fig.\ref{fig:enernew} our values
of the energy per nucleon $\varepsilon_A$ calculated for two version are denoted by solid
lines (Mass Mod.) and solutions of the ordinary Walecka model with constant mass, denoted
by dashed lines, are presented for comparison. Our EoS's are generally much softer - from
the unrealistic value of $K^{-1}=560MeV$ for the Walecka model ($S_2$) to the reasonable
$K^{-1}=290MeV$ obtained in our model. Below saturation density  these solutions are of
course identical for a given version (solid lines). Our energy and pressure results for
$S_2$ parametrization are similar to the DBHF results
Figs.(\ref{fig:enernew},\ref{main}). The EoS for ZM model seems to be too soft for high
densities. The nucleon masses: $M_N$ and $M_{med}$ in medium with their effective masses
$M^*_N$ and $M^*_{med}$  (used in Walecka and our model respectively) are compare in
Figs.\ref{fig:mass}.

Our pressure results (lower and upper panel of Fig.\ref{main}) are compared with a
semi-experimental estimate\cite{pawel} from heavy ion collisions and indeed they correct
(solid lines) Walecka  results (dashed) quite well, making the EoS significantly softer.
We have good course of EOS in NM (lower panel) for a set $S_2$ up to density
$\varrho=5fm^{-3}$. Our results are close (slightly below for lower density) DBHF results
(dotted line) which produce the EoS enable to describe\cite{NSTARS} the mass of PSR
J1614–2230 star\cite{pulsar}. In fact, for this density, the (partial) de-confinement is
expect which will change EOS above the phase transition\cite{quarkmatter}. Therefore it
is interesting how strong, in the realistic NN calculations with off-shell effects, is
violation of the longitudinal MSR. It is worth to mention that in DBHF method, there are
additional corrections\cite{bonn} from self energy which diminish the nucleon mass with
density. Our neutron matter results Fig.\ref{main} (upper panel) for $S_2$
parametrization fit well the allowed course of EoS and can be compare with another RMF
models\cite{GlenMosz,hansel}. Anyway, in case of an additional large softening of EoS the
$S_1$ parametrization, which is much stiffer but near the allowed range, can be consider.

Strangeness corrections will be present in the strange nuclear matter\cite{Kap,strange}
which supposedly exist also in the neutron stars\cite{GlenMosz}. Because the coupling of
the hyperon to the omega mesons is weaker then that of the nucleon a shift in baryon
content from nucleons to hyperons occurs only when the shift softens the equation of
state. The generalized Hugenholz van Hove theorem concerns\cite{GHvH} different barions
in a nuclear matter; for example additional $S$ strange barions. Analogously to
(\ref{ef}) a sum of all Fermi energies, including Fermi energies of strange barions
$E^S_F$, is equal:
\begin{eqnarray}
AE^A_F+SE^S_F&=&(A\!+\!S)(\varepsilon_{A+S}+p_H/\varrho) \label{hyperons}\\
~~with~~ p_H&=&\varrho^2\frac{d}{d\varrho}\left(\varepsilon_{A+S} \right)
\end{eqnarray}
Therefore, the medium corrections  to the mass of strange barions like $\Lambda$ and
$\Sigma$ based on Eq.\ref{hyperons} will be similar to (\ref{masseff}) depending
mainly from a pressure and a total energy of the system. The strong
repulsion\cite{Dabr} of the $\Sigma$ particle in medium will delay the appearance (in
increasing density) of the first hyperon $\Sigma$. The corresponding EOS with the
strangeness and nucleon resonances will be calculated and published elsewhere. The
basic conclusions however will remain the same.

Our mass corrections  reduce the violation of longitudinal MSR from $50\%$ (in the
linear Walecka model), to $10\%$ in our model$^2$ with $K^{-1}=290MeV$. Other features
of the Walecka model, including a good value of the spin-orbit force remain in our
model unchanged. The presented EoS is relatively stiff above $\varrho=5fm^{-3}$ which
is desirable in the investigation of neutron and compact stars\cite{last}. The
strangeness\cite{strange} will probably not spoil the allowed course or the phase
order transition to the quark matter might happen earlier\cite{quarkmatter}. Our
softening correction to the nucleon mass will disappear naturally with deconfinement.
However in the interacting system a part of nucleons occupy states above the Fermi
level. Therefore our formula (\ref{RMF}) and MSR should be treated as the RMF
approximation. Alternatively the mean field scenario should be supplemented by
neutron-proton short range correlations which have the remarkably similar $A$
dependence as the EMC effect\cite{pias}. On the other hand the simultaneous
description of the nuclear Drell-Yan reaction\cite{drell} and the EMC effect \cite{RW}
provides that the RMF model is working correctly.
\section{Conclusions}
The conservation of a parton MSR for a positive pressure modifies the nuclear SF and
enables\footnote{Please note that in a widely used\cite{stocker,hansel} RMF model the
good compressibility is fit by nonlinear modifications of a scalar meson field with
the help of two additional parameters..} to obtain the good compressibility of NM
without free parameters using the simple linear scalar-vector W model in the RMF
approach. Particulary, it was shown that a violation of a longitudinal MSR for partons
in compressed NM can be removed by finite volumes corrections to the nucleon mass in
medium, which reduce the nuclear stiffness to the acceptable value giving the good
course of EOS for higher densities. Also we have argued, that the rather weak
dependence of the nucleon radius (or a size of confining region) from density gives
the proper EOS fit to heavy ion collisions and neutron star properties (a mass-radius
constraint), especially the most massive known neutron star[37] recently discussed in
the application to the nuclear EoS in compact and neutron stars.

\noindent Partial support of the Ministry of Science and Higher Education, Project No.
N N202046237, is acknowledged.

\end{document}